\documentstyle[12pt]{article}

\begin{document}
\begin{titlepage}
\begin{flushright}
YCTP-N1-95\\ January, 1995
\end{flushright}
\vspace{1cm}

\begin{center}
{\bf \Large Quantum Dissipation}

\vspace{1cm}

 { Aurel BULGAC$^{a,b,}$\footnote{Internet: bulgac@phys.washington.edu},
Giu DO DANG$^{b,}$\footnote{Internet: dodang@psisun.u-psud.fr} and
Dimitri KUSNEZOV$^{c,}$\footnote{Internet: dimitri@nst.physics.yale.edu.}}\\

\vspace{5mm}

{\sl $^a$ Department of Physics, FM--15, University of
Washington, Seattle, WA 98195, USA\footnote{permanent address}}\\
\vspace{3mm}

{\sl $^b$ Laboratoire de Physique Th\'eorique et Hautes Energies,\\
      Universit\'e de Paris--Sud,  B\^at. 211, 91405 Orsay, FRANCE}\\

\vspace{3mm}

{\sl $^c$ Center for Theoretical Physics,
Sloane~Physics~Laboratory,\\ Yale University, New Haven, CT 06511-8167,USA}\\

\end{center}

\vskip 2.5 cm

\begin{abstract}

We address the question of the microscopic origin of dissipation  in
collective motion of a quantum many--body system in the framework of a
parametric random matrix approach to the intrinsic dynamics. We show
that the fluctuation--dissipation theorem is generally violated and,
moreover, energy diffusion has a markedly non--Gaussian character and
the corresponding distribution has very long tails. Such features do
not support a Langevin or Fokker--Planck approach to dissipation in
collective nuclear motion.\\
\end{abstract}

\vspace{1cm}


\vspace*{\fill}\pagebreak
\end{titlepage}
%%%%%%%%%%%%%%%%%%%%%%%%%%%%%%%%%%%%%%%%%%%%%%%%%%%%%%%%%%%%%%%%%%%%%%%%%%%%%%%
\setcounter{page}{2}
%%%%%%%%%%%%%%%%%%%%%%%%%%%%%%%%%%%%%%%%%%%%%%%%%%%%%%%%%%%%%%%%%%%%%%%%%%%%%%%

While theory and experiment have gone a long way in the study of the
collective nature of large amplitude nuclear motion, the theoretical
understanding of the coupling between the collective  and the intrinsic
degrees of freedom is still in its infancy. Most of the approaches are
more or less phenomenological in nature. Hill and Wheeler \cite{hil}
suggested almost forty years ago that Landau--Zener transitions are at
the origin of nuclear dissipation. Over the years there have been a
relatively large number of studies of this particular mechanism
\cite{str} and the range of results is equally diverse. Implicit in
this interpretation is the presumption that irreversibility is of
quantum origin. Even though there exist quantum approaches (the linear
response model \cite{sie}, the hopping model \cite{bus}, path integral
method \cite{bri} and others),  many formulations  are basically classical
, e.g. the so called ``wall formula'' \cite{wal}. These include
the more pragmatic phenomenological models, such as the Langevin
equation or Fokker--Planck equation \cite{rrr}, Maxwell's model for
friction with memory  effects \cite{nor}, and to a certain extent
kinetic approaches, e.g. two--body dissipation  mechanism \cite{wal}. In
an analysis of a generic problem of coupled slow and fast degrees of
freedom, Berry and Robbins \cite {ber} obtained friction for the slow
subsystem only by treating the entire system classically, attributing
it to ``a clash between
the essence of quantization, namely the discrete spectrum of
frequencies, and the essence of chaos, namely mixing and a continuous
spectrum extending to zero frequency'' \cite{ber}. The present status of
our understanding of the microscopic origin of dissipation in many--body
quantum systems is thus rather unsatisfactory. In this letter we explore
the nature of dissipation in a many-body system, using parametric random
matrix theory, which allows for  direct solution in many cases of the
quantum dynamics of the system.

The physical systems we explore are many-body systems which exhibit excitations
on two distinct time scales, described by collective (slow) and  intrinsic
(fast) degrees of freedom. In order to address how energy is  transferred from
the slow $(X,P)$ to the fast $(x,p)$ modes, we will assume that the slow modes
evolve classically at constant velocity $V_0$ according to $X(t)=V_0 t$.
(Although we do not consider it here, this restriction can be lifted, and the
more general problem solved using the results and methods  presented here.) As
a consequence of this assumption, we can solve for the quantum dynamics of the
fast subsystem, and even obtain analytic results for situations  which are
analogous to the conventional adiabatic and diabatic limits. The intrinsic
system is defined by its matrix elements, and is taken as  complex, described
by its average level density, $\rho(E)$, and its spectral fluctuations, in the
form
%%%%%%%%%%%%%%%%%%%%%%%%%%%%%%%%%%%%%%%%%%%%%%%%%%%%%%%%%%%%%%%%%%%%%%%%%%%%%%
\begin{equation}
H(X) = H_0 + H_1(X) .  \label{eq:hint}
\end{equation}
%%%%%%%%%%%%%%%%%%%%%%%%%%%%%%%%%%%%%%%%%%%%%%%%%%%%%%%%%%%%%%%%%%%%%%%%%%%%%%
Here $H_0$ is chosen to be a diagonal $N\times N$ matrix,  defining the average
density of states, with $\langle k|H_0|l\rangle = [H_0]_{kl}=\varepsilon _k
\delta _{kl}$. In the  basis of the eigenstates of $H_0$, we define $H_1(X)$ as
a parameter dependent, $N\times N$ real Gaussian random matrix, which is
completely specified by its first two moments
%%%%%%%%%%%%%%%%%%%%%%%%%%%%%%%%%%%%%%%%%%%%%%%%%%%%%%%%%%%%%%%%%%%%%%%%%%%%%%
\begin{eqnarray}
\overline{ [H_1(X)]_{ij}[H_1(Y)]_{kl} } &=&
[\delta _{ik}\delta _{jl}+\delta _{il}\delta_{jk}]{\cal
F}_{ij}(X-Y),\nonumber\\
\overline{[H_1(X)]_{kl}} &=&0.
\end{eqnarray}
%%%%%%%%%%%%%%%%%%%%%%%%%%%%%%%%%%%%%%%%%%%%%%%%%%%%%%%%%%%%%%%%%%%%%%%%%%%%%%
${\cal F}_{ij}(X-Y)$ is a ``bell--shaped'' correlation function with a
characteristic width $X_0$, and the overline stands for the ensemble
average. The dependence on $i,j$ allows for the description of banded
matrices, where an effective number of states $N_0\leq N$ can be coupled
by $H_{1}(X)$. Such a parametrization implies that correlations between
different instantaneous spectra corresponding to different `shapes' $X$
are effective only within a distance $\approx X_0$. The average level
density for each fixed shape $X$ is given by $H_0$, while its spectral
fluctuation properties (in this case GOE, but all our formalism applies
to GUE as well) are determined by $H_1(X)$. We use a convenient
parametrization of the correlator ${\cal F}_{ij}(X)$ introduced in Refs.
\cite{bri}
%%%%%%%%%%%%%%%%%%%%%%%%%%%%%%%%%%%%%%%%%%%%%%%%%%%%%%%%%%%%%%%%%%%%%%%%%%%
\begin{equation}
{\cal F}_{ij}(X)=\frac{W_0}{\sqrt{\rho (\varepsilon _i)\rho (\varepsilon _j)}}
\exp \left [ -\frac{(\varepsilon _i -\varepsilon _j)^2}{2\kappa _0
^2} \right ]F \left (\frac{X}{X_0}\right ).
\end{equation}
%%%%%%%%%%%%%%%%%%%%%%%%%%%%%%%%%%%%%%%%%%%%%%%%%%%%%%%%%%%%%%%%%%%%%%%%%%%
Here $F(x)=F(-x)=F^*(x)\le 1$, $F(0)=1$ and $W_0, \; \kappa _0 $ ($N_0
\approx \kappa _0 \rho(\varepsilon ))$  and $X_0$ are characteristic to
the given system. The instantaneous spectra of a Hamiltonian $H(X)$
with constant average level density, $[H_0]_{kl}=k\delta _{kl}$, is
shown in Fig. 1 for gaussian  $F(x)=\exp(-x^2/2)$ (top) and
exponential $F(x)=\exp(-|x|)$ (bottom) correlations. Notice that a conventional
adiabatic limit does not exist for the exponential, as the individual
energy levels undergo Brownian motion, and are not smooth on any time scale.

The time evolution of this system is found by solving  the
time--dependent Schr\"{o}dinger equation:
%%%%%%%%%%%%%%%%%%%%%%%%%%%%%%%%%%%%%%%%%%%%%%%%%%%%%%%%%%%%%%%%%%%%%%%%%%%%%%
\begin{equation}
\psi (t ) = {\mathrm{T}}\exp \left [-\frac{i}{\hbar } \int _0 ^{t}\! d s
H(X(s))
      \right ]\psi (0)= {\cal {U}}(t)\psi (0). \label{eq:psi}
\end{equation}
%%%%%%%%%%%%%%%%%%%%%%%%%%%%%%%%%%%%%%%%%%%%%%%%%%%%%%%%%%%%%%%%%%%%%%%%%%%%%
where $\mathrm{T}$ is the time-ordering operator, and ${\cal {U}}(t)$
the propagator. (We assume that the initial state $\psi(0)$ is
uncorrelated with the Hamiltonian $H(X(t))$ at later times.)  The
average propagator $U(t)=\overline{{\cal U}(t)}$, found by using Eqs.
(2) and resumming all leading order diagrams in perturbation expansion
of ${\cal U}(t)$ in the limit $N_0\gg 1$, satisfies the system of
integral equations [11]:
%%%%%%%%%%%%%%%%%%%%%%%%%%%%%%%%%%%%%%%%%%%%%%%%%%%%%%%%%%%%%%%%%%%%%%%%%%%%%%
\begin{eqnarray}
U_k(t )&=&U_{0\; k}(t ) - \frac{1}{\hbar ^2}\int _0^{t } \!ds_1 \!
\int _0^{s_1} \! ds_2
U_{0\; k}(s_2) U_k(t -s_1)\nonumber\\
& &\times \sum _{n=1}^{N} {\cal F}_{kn}(X(s_1)-X(s_2))U_n(s_1-s_2) ,
\end{eqnarray}
%%%%%%%%%%%%%%%%%%%%%%%%%%%%%%%%%%%%%%%%%%%%%%%%%%%%%%%%%%%%%%%%%%%%%%%%%%%%%%
where $U_{0\;k}(\tau) = \exp(-i\varepsilon _k t/\hbar )$.
In order to compute averages of observables, we introduce the set of
generalized occupation number probabilities
%%%%%%%%%%%%%%%%%%%%%%%%%%%%%%%%%%%%%%%%%%%%%%%%%%%%%%%%%%%%%%%%%%%%%%%%%%%%%%
\begin{eqnarray}
{\cal N}_{k}(t_1,t_2)&=&
\overline{\langle \psi (t_1)|k\rangle \langle k|\psi (t_2)\rangle }\nonumber\\
&=&\sum_l \overline{\langle l |{\cal {U}}^{\dagger }
(t_1)|k\rangle \langle k|{\cal U}(t_2)| l\rangle } n_l(0),
 \label{eq:tkdef}
\end{eqnarray}
%%%%%%%%%%%%%%%%%%%%%%%%%%%%%%%%%%%%%%%%%%%%%%%%%%%%%%%%%%%%%%%%%%%%%%%%%%%%%%
where $n_{l}(t)\equiv {\cal N}_{l}(t,t)$  is the occupation probability
of the state $|l\rangle$. ${\cal N}_{k}(t_1,t_2)$ satisfy the following
set of integral equations
%%%%%%%%%%%%%%%%%%%%%%%%%%%%%%%%%%%%%%%%%%%%%%%%%%%%%%%%%%%%%%%%%%%%%%%%%%%%%%
\begin{eqnarray}
{\cal N}_k(t_1,t_2) &=&  U_k^*(t_1)U_k(t_2)n_k(0)
 + \frac{1}{\hbar ^2} \int _0^{t_1}\!\! ds_1\!\! \int _0^{t_2} \!\!
 ds_2 \sum _l {\cal N}_l(s_1,s_2)
 {\cal F}_{lk}(s_1 -s_2)\nonumber\\
& &\times U_k^*(t_1 -s_1)U_k(t_2-s_2).
\end{eqnarray}
%%%%%%%%%%%%%%%%%%%%%%%%%%%%%%%%%%%%%%%%%%%%%%%%%%%%%%%%%%%%%%%%%%%%%%%%%%%%%%
These equations specify the time evolution of the system, and we will
consider $(i)$ the numerical solutions of (5)-(7) and the velocity
dependence of the diffusion constant, and $(ii)$ the  extension of the
formalism to the regime  $1\ll N_0< N=\infty$, where we find analytic
limits and a great simplification of the formalism as well.

The first situation we study is that of constant average level density
($[H_0]_{kl}=k\delta_{kl}$), as in a stadium billiard. Eqs. (5)-(7) have
been solved numerically for $N=101$ levels, a bandwidth $N_0=21$,   a
gaussian correlation $F(x)=\exp(-x^2/2)$ with $X_0=1$, and initial
conditions $n_{k}(0)=\delta_{k,51}$.  The resulting occupation numbers
$n_k(t)$ are shown in Fig. 2 for the cases of fast ($V_0=4$, top) and
slow  ($V_0=1/16$, bottom) driving velocities. As the results are
symmetric with respect to the index $k$, $n_k(t)=n_{102-k}(t)$, only
$k=1-51$ are shown, counting from the bottom of the figure. One might
expect that even a small driving velocity would result in a complicated
time evolution, as the Hamiltonian is time--dependent, and has many
small gaps in the instantaneous spectrum, where Landau--Zener
transitions might occur and thus induce ``irreversibility'' \cite{hil}.
Actually, as we have discussed at length in Ref.
\cite{bdk}, this mechanism, which has been advocated in many previous
treatments \cite{str}, is valid only for isolated level crossings and
thus is unrealistic when there are many non--isolated ones as shown in
Fig.1.

In Fig. 2, one can clearly distinguish two time scales: a relatively
rapid initial transient evolution, followed by a much slower one. While
the initial transient behaviour is almost identical in both cases,
governed by the same spreading width $\Gamma ^{\downarrow }$, the long
time behaviour is strikingly different. For small driving velocities,
the time evolution rapidly equilibrates, and can be understood in terms
of the  $V_0\rightarrow 0$ limit, corresponding to constant random
matrix theory \cite{bdk1}. For large velocities there is a steady
evolution to a different probability distribution. The initial transient
behaviour arises only because our initial occupation probabilities
$n_k(0)$ did not originate from an instantaneous eigenstate of $H(0)$ (detailed
discussion on initial conditions will be presented elsewhere\cite{bdk}).
The subsequent long time behaviour is due to the explicit time
dependence of the Hamiltonian $H(t)$ and would be absent for a time
independent one.

The diffusion process associated with these time evolutions can
be characterized by the energy variance, ${\Delta_E}(t)$, and the energy
diffusion constant $D(V_0^2)$, defined by
%%%%%%%%%%%%%%%%%%%%%%%%%%%%%%%%%%%%%%%%%%%%%%%%%%%%%%%%%%%%%%%%%%%%%%%%%%%%%%
\vspace{3mm}

\begin{picture}(450,40)
\put(30,20){${\Delta_E}(t)= \overline{ \langle \psi (t)|
[H(t)-E(t)]^2| \psi (t) \rangle }$}
\put(220,20){$\quad \approx \quad {\rm const}\; + 2D(V_0^2)t.$ }
\put(420,20){(8)}
\put(220,8){$t\rightarrow\infty$}
\end{picture}

\setcounter{equation}{8}
%%%%%%%%%%%%%%%%%%%%%%%%%%%%%%%%%%%%%%%%%%%%%%%%%%%%%%%%%%%%%%%%%%%%%%%%%%%%%%
\noindent In Fig. 3, $D(V_0^2)$ can be seen to exhibit quadratic
velocity dependence, in contradiction to previous claims \cite{str}. In
the case we consider here of a symmetrical initial distribution $n_k(0)$
and constant level density, the average energy $E(t)= \overline{
\langle\psi (t)|H(t)| \psi (t) \rangle }$ is constant, hence the
reaction force on the slow system, in particular the friction force,
exactly vanishes. This is consistent with a fluctuation--dissipation
theorem in the following sense. Expressed as $\gamma =\beta D$, where
$\gamma$ is the ``friction'' coefficient and $\beta =1/T=d \ln \rho
(e)/d e\equiv 0$ is the inverse thermodynamic temperature, we have the
expected result $dE(t)/d t = \gamma \equiv 0$.

We shall outline briefly a further extension and simplification of the
formalism  in the limit $N\rightarrow\infty$, i.e. in the spirit of the
standard constant random matrix theory, by introducing the
characteristic functional \cite{bdk1}
%%%%%%%%%%%%%%%%%%%%%%%%%%%%%%%%%%%%%%%%%%%%%%%%%%%%%%%%%%%%%%%%%%%%%%%%%%%
\begin{eqnarray}
{\cal {N}}(t_1,t_2,\tau)& = & \overline{\langle \psi (t_1)|
\exp \left [ \frac{iH_0(\tau -t_1+t_2) }{\hbar } \right ]
|\psi (t_2)\rangle }, \nonumber\\
{\cal {N}}_k(t_1,t_2)& = & \frac{1}{2\pi \hbar \rho (\varepsilon _k)}
\int  d \tau {\cal {N}}(t_1,t_2,\tau)\nonumber\\
& &\times \exp \left [ -\frac{i\varepsilon _k(\tau -t_1+t_2) }{\hbar }
 \right ] , \\
{\cal {N}}(t,t,\tau )& = &
\exp \left [\sum _n
\overline{\langle \!\langle \psi (t) |H_0^n |\psi (t)\rangle \!\rangle }
\frac{(i\tau )^n}{\hbar ^n n ! } \right ],
\end{eqnarray}
%%%%%%%%%%%%%%%%%%%%%%%%%%%%%%%%%%%%%%%%%%%%%%%%%%%%%%%%%%%%%%%%%%%%%%%%%%%
where $\overline{\langle \!\langle \psi (t)| H_0^n |\psi (t) \rangle
\!\rangle} $ are cumulants. ${\cal {N}}(t_1,t_2,\tau)$ satisfies the
evolution equation
%%%%%%%%%%%%%%%%%%%%%%%%%%%%%%%%%%%%%%%%%%%%%%%%%%%%%%%%%%%%%%%%%%%%%%%%%%%
\begin{eqnarray}
{\cal {N}}(t_1,t_2,\tau )&=& \sigma ^*(t_1)\sigma (t_2)+
\frac{\sqrt{2\pi }\kappa _0 W_0}{\hbar^2 }
\int _0^{t_1}\! d s_1\! \int _0^{t_2}\! d s_2 \nonumber\\
 & &\times {\cal {N}}(s_1 ,s_2 ,\tau )
  \exp \left [ \frac{\kappa _0 ^2}{2}
\left ( \frac{\beta }{2}+i\frac{s_1 - s_2 -\tau}{\hbar}
\right ) ^2 \right ] \nonumber\\
& &\times F \left ( \frac{(s_1 -s_2 )V_0}{X_0}\right )
\sigma ^*(t_1-s_1)\sigma(t_2-s_2),
\end{eqnarray}
%%%%%%%%%%%%%%%%%%%%%%%%%%%%%%%%%%%%%%%%%%%%%%%%%%%%%%%%%%%%%%%%%%%%%%%%%%%
where $\sigma (t)=\exp (i\varepsilon _k t/\hbar )U_k (t)$  (note $\sigma
(t)$ is state independent), for which an equation similar to Eq. (5) can
be derived. In this case, there is only one equation to be solved,
instead of $N$--coupled equations (cf. Eq. (7)), which results in a
significant simplification of the entire formalism. Moreover, various
analytic solutions can be obtained, as we exemplify below, by analyzing
the adiabatic and the diabatic evolutions of the occupation numbers for
a system with a realistic level density  of the form $\rho(\varepsilon
)=\rho _0 \exp (\beta \varepsilon)$.  $\beta =0$ corresponds to the case
we have just described, of constant average level density, while the
case of finite $\beta$ approximates fairly well a many--fermion system.

The {\it adiabatic limit} corresponds to $\kappa _0 X_0/\hbar V_0\gg 1$
(and also $\kappa _0 \beta \ll 1 $), from which we find
%%%%%%%%%%%%%%%%%%%%%%%%%%%%%%%%%%%%%%%%%%%%%%%%%%%%%%%%%%%%%%%%%%%%%%%%%%%
\begin{equation}
{\cal {N}}(t,t,\tau)=
\exp
\left \{ \frac{2\pi W_0}{\hbar }
\left [ F\left (\frac{\tau V_0 }{ X_0}\right ) -1
\right ] t-\frac{2\pi W_0|\tau |}{\hbar }\right \} .
\end{equation}
%%%%%%%%%%%%%%%%%%%%%%%%%%%%%%%%%%%%%%%%%%%%%%%%%%%%%%%%%%%%%%%%%%%%%%%%%%%
All odd moments of $H_0$ vanish identically (since $F(x)=F(-x)$), and in
the limit $t \rightarrow \infty $, all even cumulants of $H_0$ increase
linearly in time. If $F(x)=\exp (-x^2/2)$ (we shall use this form
hereafter for illustrative purposes) then
%%%%%%%%%%%%%%%%%%%%%%%%%%%%%%%%%%%%%%%%%%%%%%%%%%%%%%%%%%%%%%%%%%%%%%%%%%%
\begin{eqnarray}
\overline{\langle \!\langle \psi (t)|H_0^{2n}|\psi (t)\rangle \!\rangle }&=&
\frac{2\pi W_0 t }{\hbar } \left (\frac{\hbar V_0}{X_0} \right ) ^{2n}
\frac{(2n)!}{2^n n!},\nonumber\\
D(V_0^2)&=&\frac{\pi \hbar W_0 V_0^2}{X_0^2}
=\frac{\hbar \Gamma ^{\downarrow} V_0^2}{2X_0^2},
\end{eqnarray}
%%%%%%%%%%%%%%%%%%%%%%%%%%%%%%%%%%%%%%%%%%%%%%%%%%%%%%%%%%%%%%%%%%%%%%%%%%%
($\Gamma ^{\downarrow}=2\pi W_0$) resulting in a non-Gaussian distribution.

In the {\it diabatic limit}, $\kappa _0 X_0/\hbar V_0 \ll 1 $,  we find
%%%%%%%%%%%%%%%%%%%%%%%%%%%%%%%%%%%%%%%%%%%%%%%%%%%%%%%%%%%%%%%%%%%%%%%%%%%
\begin{equation}
{\cal {N}}(t,t,\tau )=\exp \left \{ \frac{2\pi X_0 W_0\kappa _0 }{\hbar ^2 V_0}
\left [
\exp \left ( \frac{\kappa _0 ^2}{2}\left (\frac{\beta }{2}+\frac{i\tau }{\hbar}
 \right ) ^2 \right )
 - \exp \left( \frac{\kappa _0 ^2\beta ^2 }{8}\right ) \right ] t \right \}.
\end{equation}
%%%%%%%%%%%%%%%%%%%%%%%%%%%%%%%%%%%%%%%%%%%%%%%%%%%%%%%%%%%%%%%%%%%%%%%%%%%
In this case again all the cumulants of $H_0$ increase linearly in time
%%%%%%%%%%%%%%%%%%%%%%%%%%%%%%%%%%%%%%%%%%%%%%%%%%%%%%%%%%%%%%%%%%%%%%%%%%%
\begin{equation}
\overline{\langle \!\langle \psi (t)|H_0^n|\psi (t)\rangle \!\rangle }
=\left [ \frac{2\pi X_0 W_0 \kappa _0 }{\hbar ^2V_0}
\exp \left ( \frac{\beta ^2\kappa _0 ^2}{8}\right )
\left ( \frac{i\kappa _0 }{\sqrt{2}} \right )^n
{\rm H}_n\left ( -\frac{i\kappa _0 \beta }{2\sqrt{2}} \right ) \right ]\; t,
\end{equation}
%%%%%%%%%%%%%%%%%%%%%%%%%%%%%%%%%%%%%%%%%%%%%%%%%%%%%%%%%%%%%%%%%%%%%%%%%%%
where ${\rm H}_n(x)$ are Hermite polynomials. From the explicit
expressions for the first and second cumulants we thus obtain that
%%%%%%%%%%%%%%%%%%%%%%%%%%%%%%%%%%%%%%%%%%%%%%%%%%%%%%%%%%%%%%%%%%%%%%%%%%%
\begin{equation}
\beta D= \gamma \left ( 1+\frac{\beta ^2\kappa _0 ^2}{4}\right ),
\end{equation}
%%%%%%%%%%%%%%%%%%%%%%%%%%%%%%%%%%%%%%%%%%%%%%%%%%%%%%%%%%%%%%%%%%%%%%%%%%%
which shows that the Einstein fluctuation--dissipation theorem is
generally violated.

The most salient feature of the solutions (12) and (14)  becomes
evident when one considers the asymptotic behaviour of the cummulants.
Since cumulants of higher than second order are nonvanishing, Gaussian
processes are not obtained in any of these limiting cases for the energy
diffusion. $\langle \!\langle \psi (t)|H_0^n|\psi (t)\rangle \!\rangle $
either increase indefinitely with $n$ or increase subsequently after an
initial decrease, depending on the values of parameters. As a  result
the energy distribution has very long tails. In particular for
$V_0\equiv 0$ the distribution corresponding to Eq. (12) has a
Lorentzian shape. These features imply that a Langevin or Fokker--Planck
approach to energy dissipation is at least questionable. As we have
discussed in Ref. \cite{bdk} the present results apply equally to the GUE
case.

The present approach treats the fast subsystem quantum mechanically and
the slow subsystem classically, as has been done often in the past
\cite{str}. The energy diffusion process is described in terms of some
intrinsic characteristics of the many--body system (thermodynamic
temperature, spreading width $\Gamma ^{\downarrow}$, $\kappa _0 $ and
$X_0$) and $V_0$. It is not clear yet whether these characteristics have
a meaningful classical limit separately or only in a given combination,
and this seemingly points to an apparent lack of a classical limit for
the fast degrees of freedom ($\hbar \rightarrow 0$) of the solutions
(13), (15). In Ref. \cite{ber}, friction was obtained only in a
classical treatment of both fast and slow system, while in Ref.
\cite{bri}, dissipation and friction appear only in a explicit quantum
treatment (path integral) of the entire system and the presence of
quantum fluctuations in the slow subsystem was essential. The wall
formula \cite{wal} leads to a diffusion constant $D\propto V_0^2$ as we
have obtained here (see Fig. 3 and Eq. (13)), but is essentially a
classical result, which does not depend in any significant way on
$\hbar$, and apparently reflects a different underlying mechanism. It
will be highly desirable to identify the classical limit of the present
approach.

In conclusion, we have presented numerical and analytical solutions of the time
dependent evolution of a driven complex quantum system, such as a nucleus,
under the assumption that the number of levels is large. The parametric random
matrix approach chosen here incorporates the essential attributes of the
intrinsic dynamics, namely: an exponentially increasing level density, GOE
spectral fluctuations and loss of correlations during large amplitude
collective motion. We have shown that the resulting energy diffusion process is
highly non--Gaussian in character, that the energy distribution has very long
tails and also that the energy diffusion constant is proportional to the square
of the collective velocity.

We thank H.A. Weidenm\"uller for stimulating discussions and
correspondence, the Centre National de la Recherche Scientifique
(C.N.R.S.) and  Universit\'e de  Paris-Sud for financial support, and
the computer facilities at NERSC. This work was also partially supported
by NSF and DOE. The Laboratoire de Physique Th\'eorique et Hautes
Energies is a Laboratoire associ\'e au C.N.R.S., URA 0063.

\newpage

\begin{figure}
\vspace{12cm}
\caption{ Instantaneous eigenvalue spectrum $E_n(X)$ as a function of the
``shape'' ($X$), for a Hamiltonian of the ensemble defined by Eqs. (1--3), with
$[H_0]_{jk}=k\delta_{jk}$, using the correlator $F(x)=\exp(-x^2/2)$ (top) and
$F(x)=\exp(-|x|)$ (bottom).}

\end{figure}
\newpage
\begin{figure}
\vspace{12cm}

\caption{The time dependence of the occupation probabilities $n_k(t)$, for
$k=1,\dots , 51$ (in this case $n_k(t)=n_{102-k}(t)$), where $k$ counts from
bottom to top in the figure, for the case of fast, $V_0=4$ (top), and slow,
$V_0=1/16$ (bottom), driving velocities.}

\end{figure}
\newpage
\begin{figure}
\vspace{12cm}

\caption{The time dependence of the energy variance $\Delta_E(t)$ for a
range of velocities $V_0=4,\;2,\;1,\;0.5,\;0.25,\;0.125$ and $0.0625$.
The highest curve corresponds to the larger velocity. The insert shows
the diffusion constant $D(V_0^2)$ as a function of $V_0^2$, indicating
$D\propto V_0^2$.}
\end{figure}

\end{document}